\documentclass{desyproc}
\usepackage{graphicx}

\begin{document}

\title{Measurement of the leptonic \ensuremath{t\bar{t}} charge asymmetry in the dilepton channel with the D0 detector}

\author{{\slshape Antoine Chapelain}  for the D0 Collaboration\\[1ex]
Irfu/SPP, CEA-Saclay, France }

% please do not modify the following 5 lines
\contribID{xy}  % will be entered by the editors
\confID{7095}
\desyproc{DESY-PROC-2013-XY}
\acronym{TOP2013}
\doi            % will be entered by the editors
\maketitle

\section{Introduction}

At next-to-leading order, quantum chromodynamics predicts a \ensuremath{t\bar{t}} pair production asymmetry.
The top quark is predicted to be emitted preferentially in the direction of the incoming quark,
while the top antiquark in the direction of the incoming antiquark.
The magnitude of the asymmetry has been computed to be around $9\%$ for proton-antiproton collisions at Tevatron within the SM~\cite{BS} including electroweak corrections.
In 2011 results from CDF and D0~\cite{CDF_51_ttbar, D0_54_ttbar} have driven a
lot of attention because some of the measured asymmetries were significantly higher than the predictions based on the SM. 

In this note, we report a new measurement~\cite{D0_97_dilep} accepted by PRD of the \ensuremath{t\bar{t}} asymmetry based on leptons
produced in \ensuremath{t\bar{t}} events in the dilepton channel with the full dataset collected by the D0\ Collaboration in Run~II of the Tevatron
corresponding to an integrated luminosity of 9.7~fb$^{-1}$,
and we compare our results with the most recent predictions based on the standard model~\cite{BS}.
We use the two observables $q\times\eta$ and $\Delta\eta$, where
$q$ and $\eta$ are the charge and pseudorapidity of the lepton, and
$\Delta\eta = \eta_{\ell^+} -\eta_{\ell^-}$ is the difference in lepton pseudorapidities.
The pseudorapidity $\eta$ is defined as a function of the polar angle $\theta$ with
respect to the proton beam as \mbox{$\eta =-\ln(\tan {{\theta}\over{2}})$}. Positive (negative)
$\eta$ corresponds to a particle produced in the direction of the incoming proton (antiproton).
The single-lepton asymmetry \ensuremath{A^{\ell}_{FB}} and dilepton asymmetry \ensuremath{A^{\ell\ell}}\ are defined as
\begin{equation}
\ensuremath{\ensuremath{A^{\ell}_{\rm FB}} = \frac{N(q \times \eta>0) - N(q \times \eta<0)}{N(q\times \eta>0) + N(q \times \eta<0)}}, ~~
\ensuremath{\ensuremath{A^{\ell\ell}} = \frac{N(\Delta\eta > 0) - N(\Delta\eta < 0)}{N(\Delta\eta > 0) + N(\Delta\eta < 0)}}
\label{eq:al}
\end{equation}
where $N$ corresponds to the number of leptons satisfying a given set of selection criteria.

\section{Simulation and backgrounds}

Monte Carlo (MC) events are processed through a \textsc{geant}-based~\cite{geant}
simulation of the D0 detector.
\ensuremath{t\bar{t}} events are generated with the NLO generator \textsc{mc@nlo}~\cite{mcatnlo}. 
Electroweak backgrounds such as Drell-Yan process associated with jets and diboson production are simulated using \textsc{alpgen}~\cite{alpgen} 
and \textsc{pythia}~\cite{pythia} respectively.
The so-called instrumental background arises mainly from multijets and W+jets events in which one or two jets is misidentified as electrons or
where muons or electrons originating from the semileptonic decay of a heavy flavor hadron appear isolated.
This instrumental background is estimated directly in the data by the mean of the ``matrix method''.

\section{Event selection}

The selection of events follows the approach developed for the measurement of the \ensuremath{t\bar{t}} cross section in the dilepton channel at D0~\cite{D0_54_xs_dilep}.
We require at least two high \mbox{$p_T$}\ isolated leptons and missing energy due to the two neutrinos escaping the detector.
We define three channels requiring at least two jets: dielectron channel ($ee$) with two electrons,
electron-muon channel ($e\mu$) with one electron and one muon, and dimuon channel ($\mu\mu$) with two muons.
We define an additional channel requiring exactly one jet, one electron and one muon ($e\mu$ 1 jet).
The final selection is performed in two dimensions using informations from the $b$-quark identification and the topological
variables such as $H_{T}=p_{T}^{lepton} + \sum_{i=1}^2 p_{T}^{jet}$ or the significance in missing transverse energy \mbox{${\cal S}(\mbox{$\not\!\!E_T$})$}.
The numbers of predicted background events, as well as the expected numbers of signal
events, in the four channels are given in Table~\ref{tab:yield} and show high signal purity of the selected sample.
\begin{table}[hbt]
\renewcommand{\arraystretch}{1.8}
\caption{Numbers of total expected ($N_{\rm expected}$) and observed ($N_{\rm observed}$) events
from backgrounds and \ensuremath{t\bar{t}} signal.
Expected numbers of events are shown with their statistical uncertainties.
\label{tab:yield}}
\begin{center}
\begin{tabular}[t]{|lcccccc|}
\hline
& $Z\to \ell\ell$  & Dibosons& \parbox{2.2cm}{Multijet~and $W$+jets}
& \parbox{2.2cm}{$t\bar{t}\to \ell\ell jj$} & \parbox{1.6cm}{$N_{\rm expected}$}
& \parbox{1.3cm}{$N_{\rm observed}$} \\  \hline
$ee$ & $17.2^{+0.6}_{-0.6}$  & $2.4^{+0.1}_{-0.1}$  & $\phantom{0}4.7^{+0.4}_{-0.4}$  & $127.8^{-1.4}_{-1.4}$  & $152.1^{+1.6}_{-1.6}$  & 147 \\
$e\mu$ 2 jets & $13.7^{+0.5}_{-0.5}$ & $3.9^{+0.2}_{-0.2}$  & $16.3^{+4.0}_{-4.0}$  & $314.7^{+1.1}_{-1.1}$  & $348.6^{+4.2}_{-4.2}$  & 343 \\ 
$e\mu$ 1 jet & $\phantom{0}8.7^{+0.6}_{-0.6}$ & $3.4^{+0.2}_{-0.2}$  & $\phantom{0}2.9^{+1.7}_{-1.7}$  & $\phantom{0}61.7^{+0.5}_{-0.5}$  & $\phantom{0}76.7^{+1.9}_{-1.9}$  & \phantom{0}78 \\ 
$\mu\mu$ & $17.5^{+0.6}_{-0.6}$  & $1.9^{+0.1}_{-0.1}$ &  $\phantom{0}0.0^{+0.0}_{-0.0}$  & $\phantom{0}97.7^{+0.6}_{-0.6}$  & $117.1^{+0.8}_{-0.8}$  & 114 \\
\hline
\end{tabular}
\end{center}
\end{table}

\section{Measurements}
\begin{figure}[hbt]
\begin{center}
\includegraphics[width=.35\textwidth]{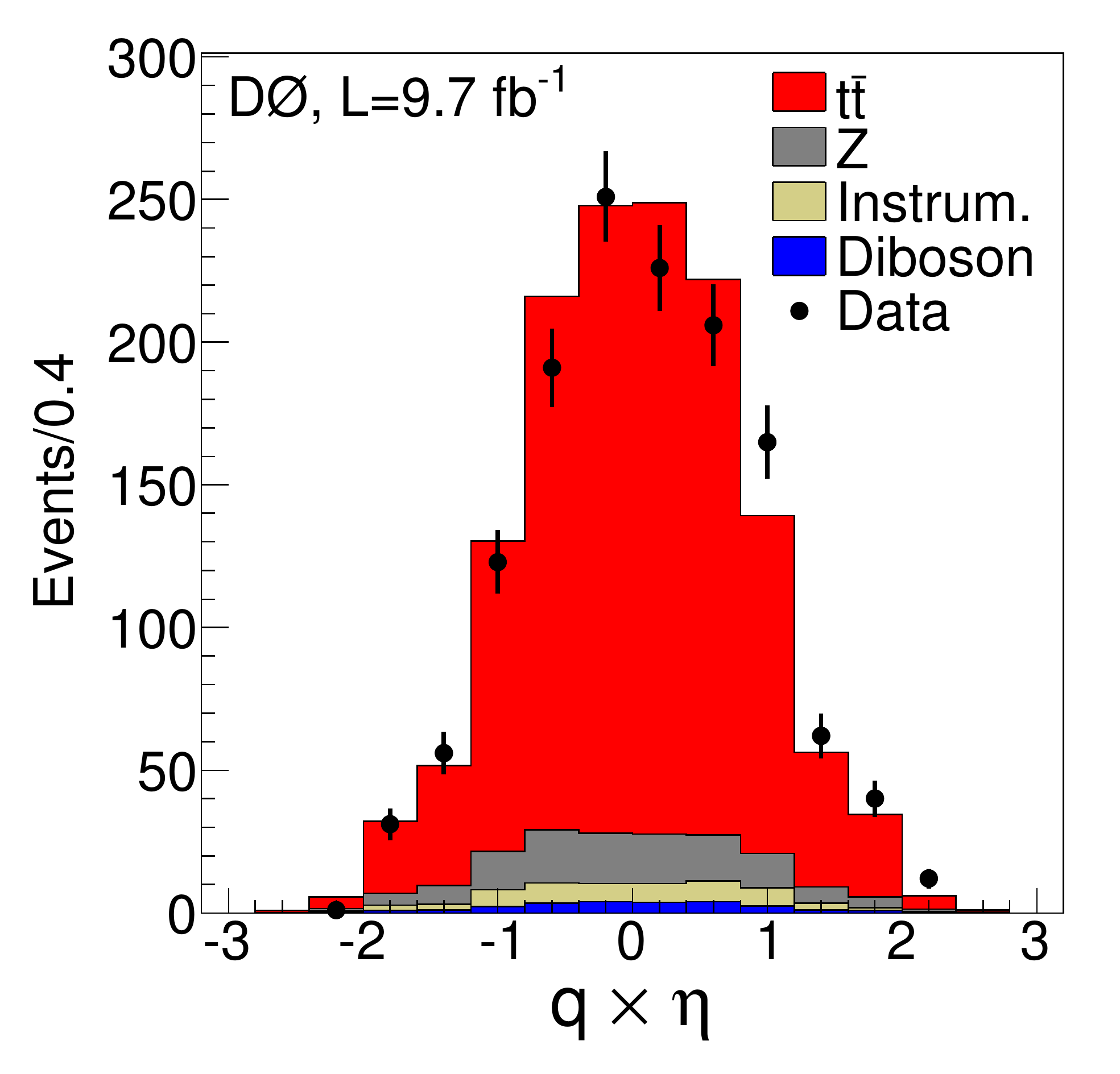}
\includegraphics[width=.35\textwidth]{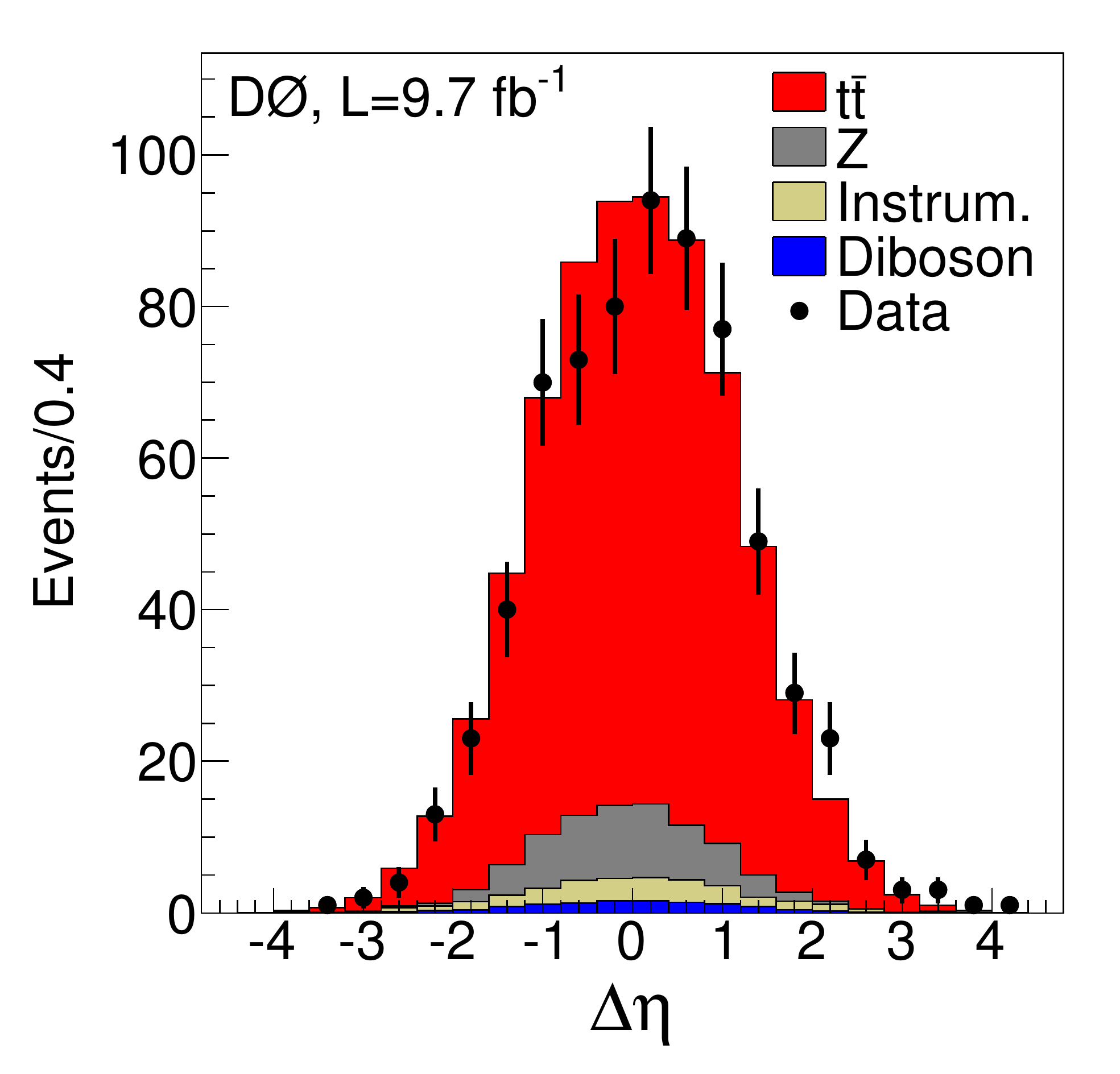}
\end{center}
\caption{Distributions in (left) $q\times\eta$ and (right) $\Delta\eta = \eta_{\ell^+} -\eta_{\ell^-}$, for the
sum of $ee$, $e\mu$ and $\mu\mu$ channels, along with predictions of the backgrounds and \ensuremath{t\bar{t}} signal.
The black points show data events and the error bars indicate the statistical uncertainty on the data.
\label{fig:reco}}
\end{figure}

\begin{figure}[hbt]
\begin{center}
\includegraphics[width=.40\textwidth]{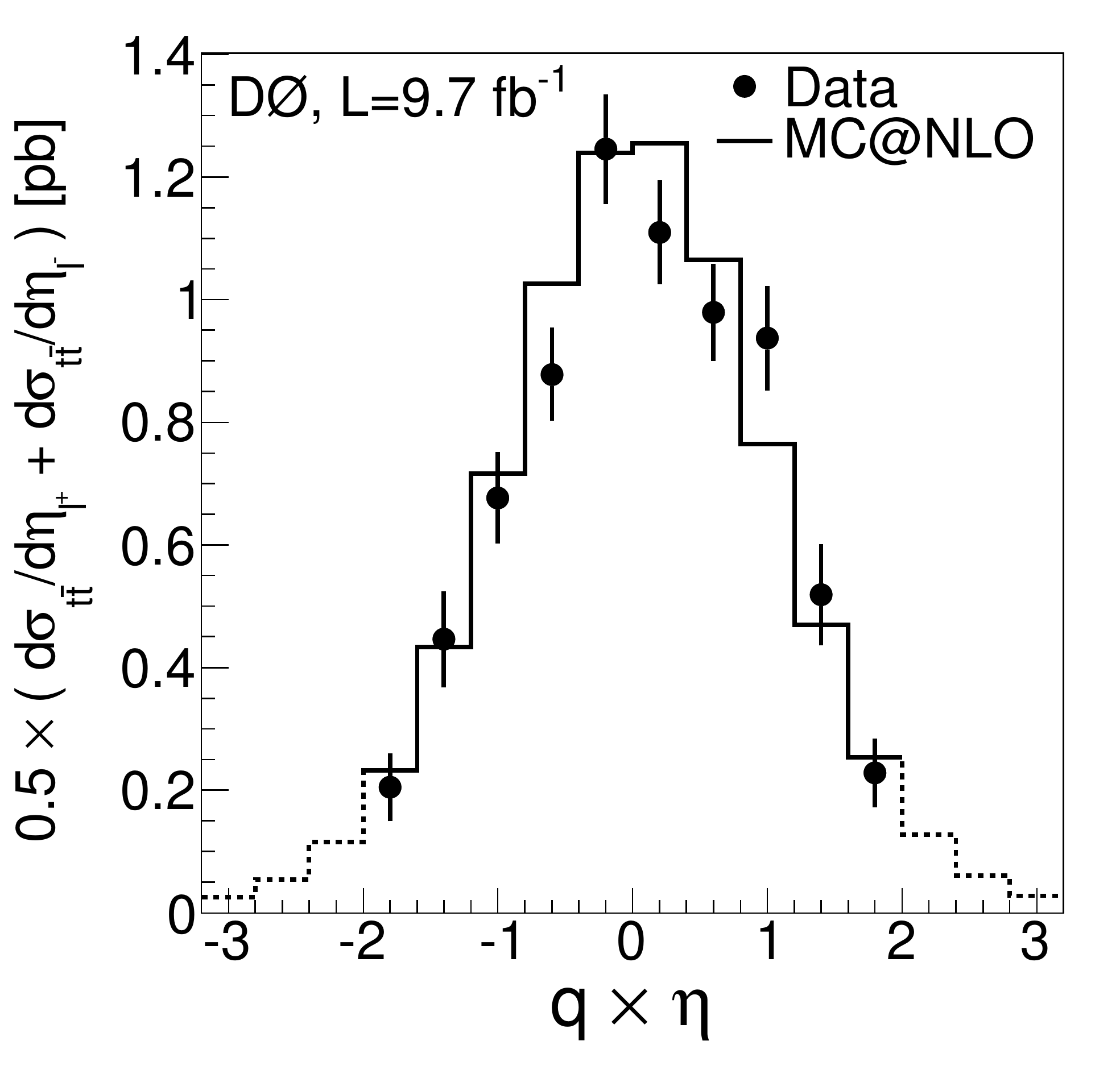}
\includegraphics[width=.40\textwidth]{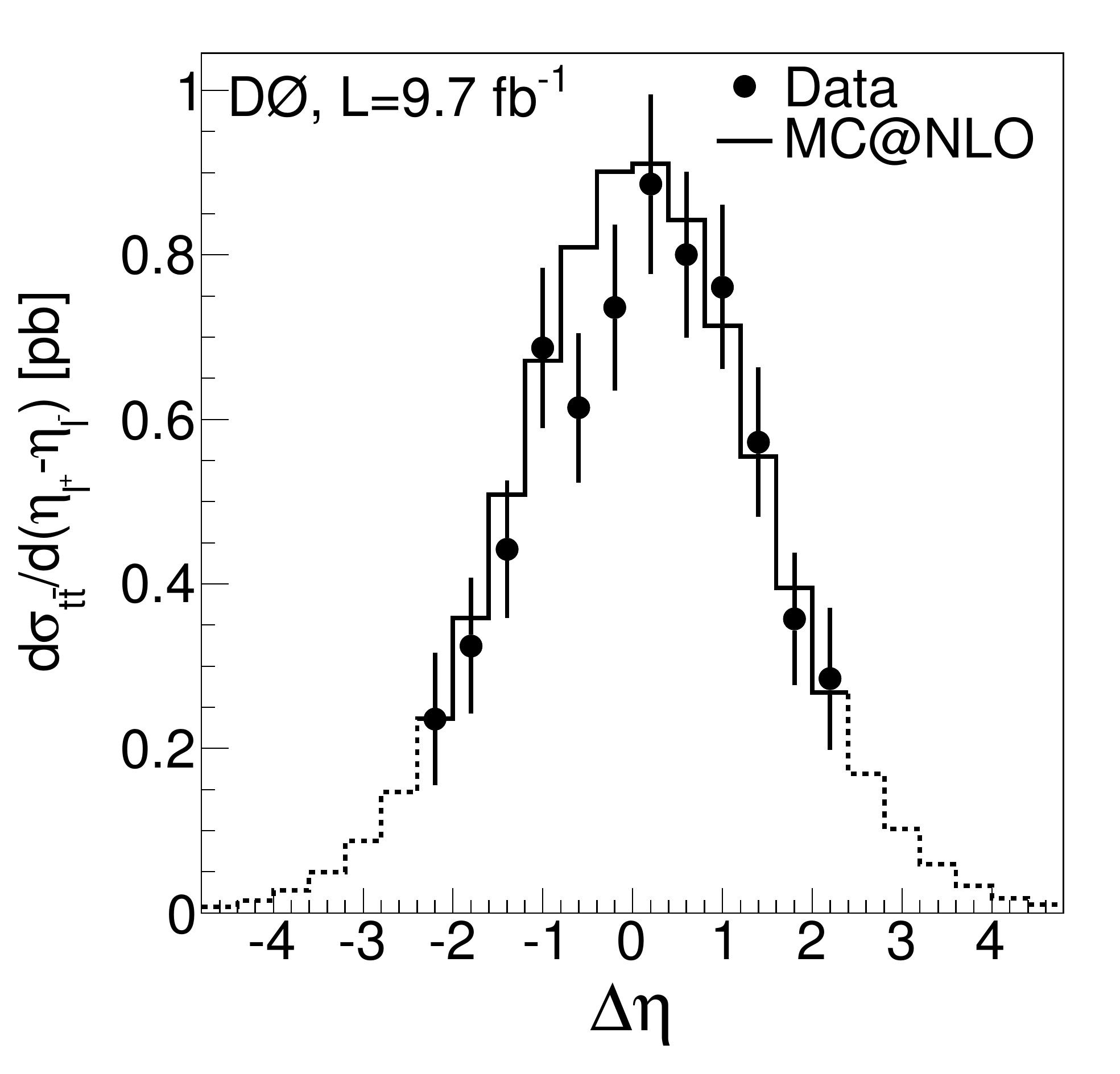} 
\end{center}
\caption{Distributions in (left) $q\times\eta$ and (right) $\Delta\eta$,
for the combined $ee$, $e\mu$, and $\mu\mu$ channels after subtraction of background and correction for selection efficiency
within the acceptance.
The error bars indicate the statistical uncertainty on data.
The dashed lines show the predictions from \textsc{mc@nlo}\ outside the analysis acceptance.
\label{fig:corr_level}}
\end{figure}

Figure~\ref{fig:reco} presents the $q\times\eta$ and $\Delta\eta$ distributions for dilepton events after applying the event selection.
To measure \ensuremath{A^{\ell}_{FB}} and \ensuremath{A^{\ell\ell}}\  we restrict the distributions to the so-called visible phase space.
This region is defined such as the statistical uncertainty on the asymmetry within the full phase space is minimized using
ensemble of pseudo datasets:  \mbox{$|\eta|<2.0$} and \mbox{$|\Delta\eta|<2.4$}.
Within each of the four channels we perform a bin-by-bin subtraction of the estimated  background contributions to the data.
We then correct bin-by-bin the background subtracted distribution for the selection efficiency to get back to the production level result
using \textsc{mc@nlo}\ \ensuremath{t\bar{t}} sample.
Figure~\ref{fig:corr_level} shows the corrected distributions for data compared to the predictions from \textsc{mc@nlo}.
Finally, we extrapolate the measured production asymmetries from the visible phase space to the full phase space 
by multiplying the asymmetries within the visible phase space with the so-called extrapolation factor.
We compute this extrapolation factor by taking the ratio of
the generator level SM \ensuremath{t\bar{t}} asymmetries from \textsc{mc@nlo}\, without selections to
asymmetries within the visible phase space.

\section{Uncertainties}

The main sources of systematic uncertainties are related to the modeling of the background and the signal
as well as instrumental uncertainty such as energy scale of jets and leptons.
The overall systematic uncertainties on $A^{\ell}_{FB}$ and $A^{\ell\ell}$ are small
compared to the statistical uncertainties (see Sec. \ref{sec:results}).
Further details about each category of uncertainty may be found in~\cite{D0_97_dilep}.

\begin{table}[h]
\caption{The measured asymmetries defined in Eq.~(\ref{eq:al})
for all channels combined within the visible and full phase spaces, compared to the predicted SM NLO asymmetries~\cite{BS} for inclusive \ensuremath{t\bar{t}} production.
The first uncertainty on the measured values corresponds to the statistical and the second to the systematic
contribution. All values are given in \%.
\label{tab:results}}
\renewcommand{\arraystretch}{1.2}
\begin{center}
\begin{tabular}{|lccc|}
\hline
  & Visible phase space & Full phase space & Prediction \\
\hline
\ensuremath{A^{\ell}_{\rm FB}(\%)} & \phantom{$-$}4.1 $\pm$ \phantom{1}3.5 $\pm$ 1.0 & \phantom{0}4.4 $\pm$ 3.7 $\pm$ 1.1 & 3.8 $\pm$ 0.3 \\
\ensuremath{A^{\ell\ell}(\%)} & \phantom{-}10.5 $\pm$ \phantom{1}4.7 $\pm$ 1.1 & 12.3 $\pm$ 5.4 $\pm$ 1.5 & 4.8 $\pm$ 0.4 \\
\hline
\end{tabular}
\end{center}
\end{table}

\section{Results}
\label{sec:results}

We combine the four channels taking into account the correlations of the different systematic uncertainties using
the BLUE method~\cite{BLUE_1, BLUE_2}.
Table~\ref{tab:results} shows the combined results within the visible and the full phase space as well as  the more recent predictions
based on the standard model~\cite{BS}.
The measured \ensuremath{A^{\ell}_{FB}} and \ensuremath{A^{\ell\ell}}\ within the full phase space are consistent with the predictions.

\begin{figure}[hbt]
\begin{center}
\includegraphics[width=.42\textwidth]{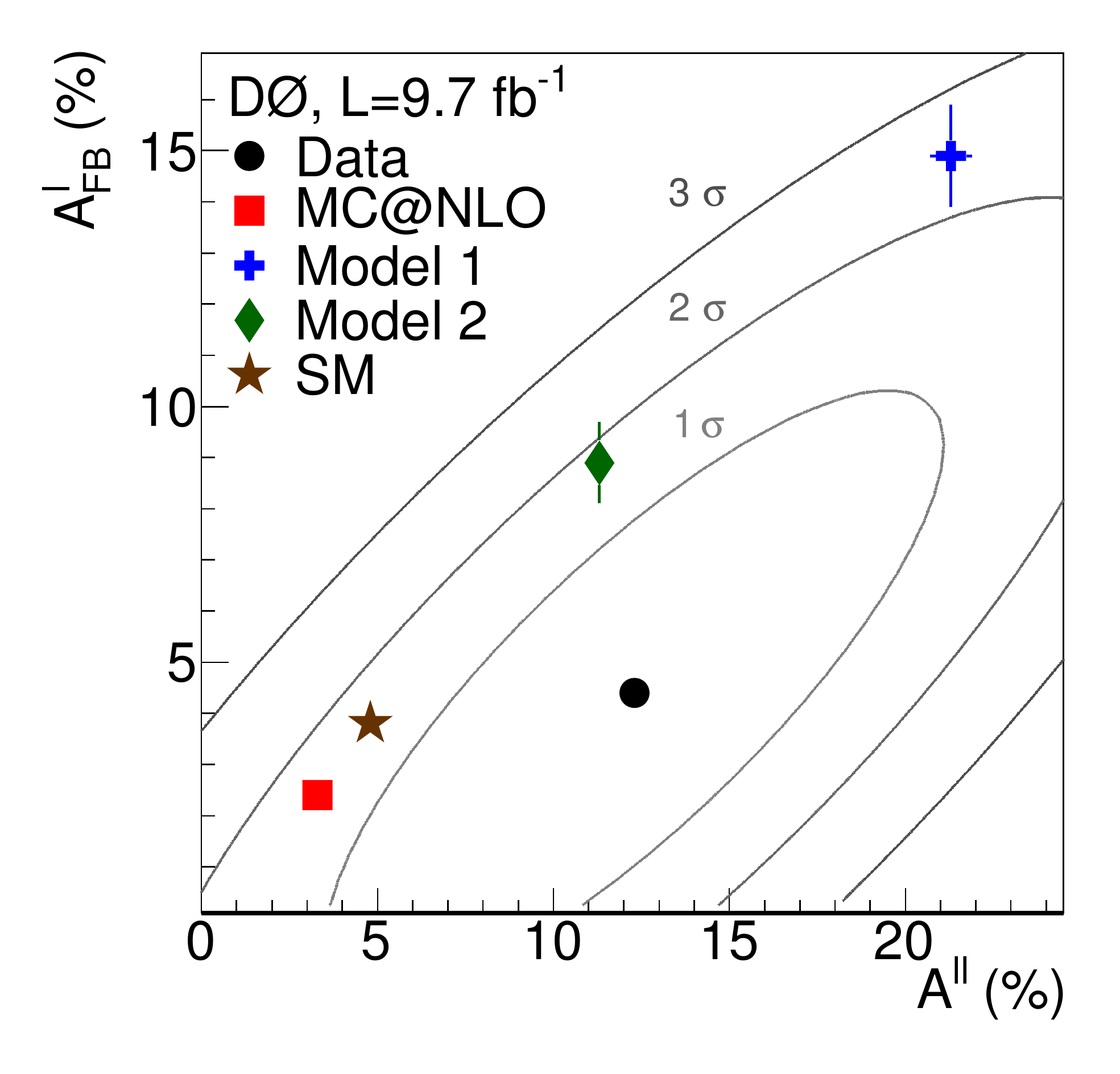}
\includegraphics[width=.42\textwidth]{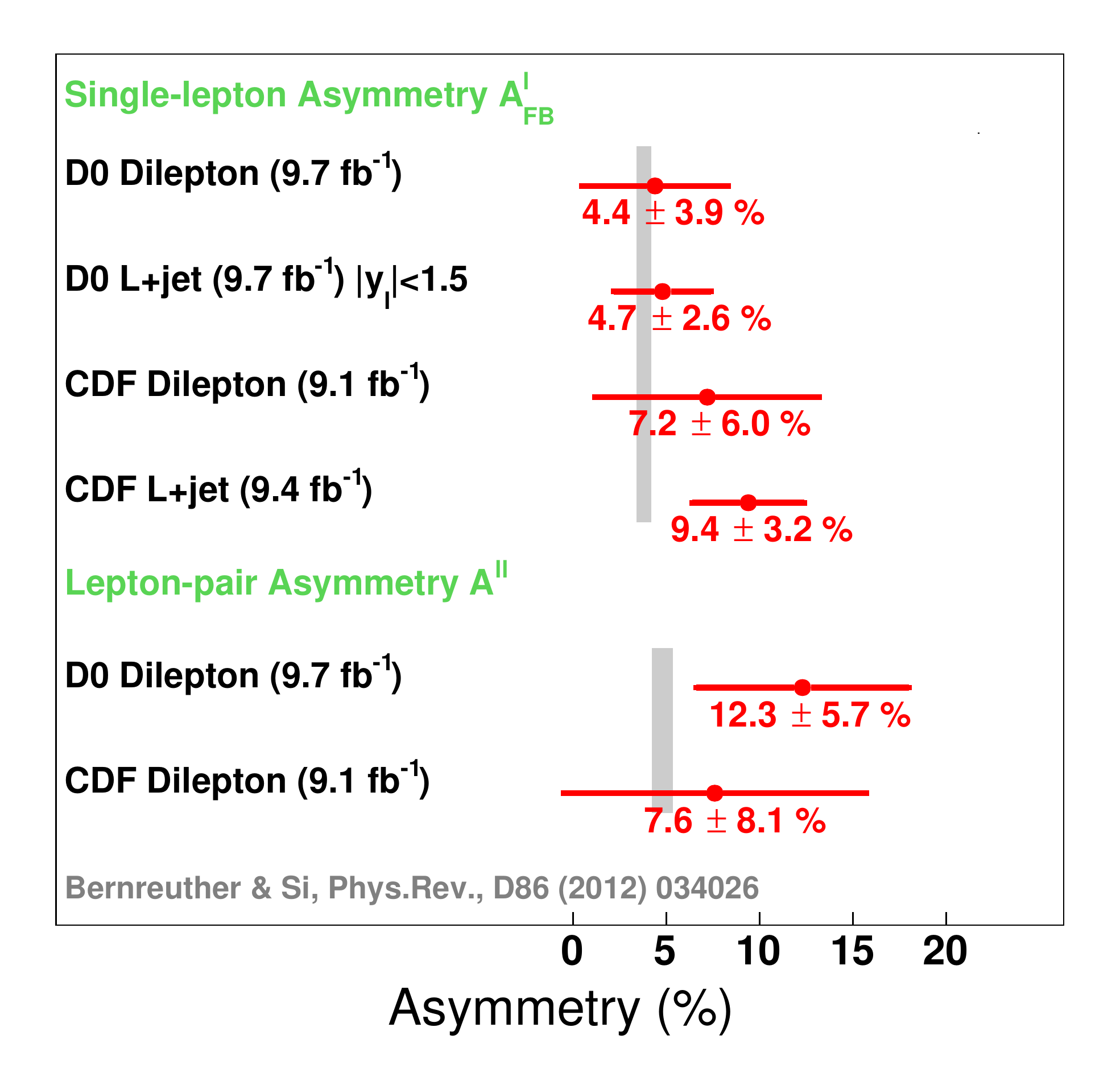}
\end{center}
\caption{(left) D0 dilepton extrapolated \ensuremath{A^{\ell}_{FB}} versus \ensuremath{A^{\ell\ell}}\ asymmetries in \ensuremath{t\bar{t}} data, the predictions from \textsc{mc@nlo},  axigluon models,
and from the latest SM NLO prediction~\cite{BS}. The ellipses represent contours of total uncertainty
at 1, 2, and 3 SD on the measured result. Predicted asymmetries are shown with their statistical uncertainties.The predictions from MC@NLO
differ from the SM ones because MC@NLO does not include the electroweak corrections.
(right) Summary of the \ensuremath{A^{\ell}_{FB}} and \ensuremath{A^{\ell\ell}} measurements at the Tevatron.
\label{fig:al_vs_all}}
\end{figure}

We measure the statistical correlation between \ensuremath{A^{\ell}_{FB}} and \ensuremath{A^{\ell\ell}}\ to be of 0.82 as explained in~\cite{D0_97_dilep} in order to compute the ratio
of these two asymmetries which allow to achieve a better sensitivity with ratio to the individual asymmetries due
to systematic uncertainties cancellation. We measure a ratio equal to $0.36 \pm 0.20$ consistent at the level of 2 standard deviations with the prediction of $0.79 \pm 0.10$.
This predicted ratio is found to be almost the same for the different tested models as can be seen in Fig.~\ref{fig:al_vs_all}(left).

CDF and D0\ both recently measured \ensuremath{A^{\ell}_{FB}} in the lepton+jets ($\ell$+jets) final state~\cite{CDF_97_ljets,D0_97_ljets} to be
$(9.4^{+3.2}_{-2.9})\%$ and  $(4.7^{+2.6}_{-2.7})\%$, respectively.
CDF also reported measurements of \ensuremath{A^{\ell}_{FB}} and \ensuremath{A^{\ell\ell}} in the dilepton final state~\cite{CDF_97_dilep} to
be $(7.2 \pm 6.0)\%$ and $(7.6 \pm 8.1)\%$, respectively.
We are able to compare our measurements performed in the dilepton channel at D0 with the results in the dilepton and $\ell$+jets at CDF since they
all are extrapolated to the full phase space. The measured \ensuremath{A^{\ell}_{FB}} on one hand and \ensuremath{A^{\ell\ell}} (dilepton only) on the other hand are in agreement.

The \ensuremath{A^{\ell}_{FB}} measurement in the $\ell$+jets channel at D0 is restricted to the region $|\eta_{lepton}|<1.5$ 
and not extrapolated to the full phase space. We cannot then compare directly with our \ensuremath{A^{\ell}_{FB}} measurement.
Nevertheless in the dilepton channel at D0 we found that the ratio of \ensuremath{A^{\ell}_{FB}} measured within the full 
and visible phase space ($|\eta_{lepton}|<2.0$) is at the order of $\sim 1.1$.
The small extrapolation correction allows to compare the two D0 \ensuremath{A^{\ell}_{FB}} results which we observe to be in agreement.
Figure~\ref{fig:al_vs_all}(right) shows a summary of the Tevatron measurement.

The combination of the CDF and D0\ results will be the last step to build the legacy measurement from the Tevatron.
We can perform the combination in different ways. One of them could be to combine and extrapolate the measurements at the same time.
Using the distributions of the asymmetry as a function of the lepton pseudorapidity from each CDF and D0 measurements, we can fit them separately and then combine
the parameters of the fit function using the BLUE method. 
It is then straightforward to extract the asymmetry in the full phase space (as well as in any restricted phase-space regions).
The statistical uncertainty of the combined measurements are expected to be $\sim1.5\%$ for \ensuremath{A^{\ell}_{FB}} and $\sim4.6\%$ for \ensuremath{A^{\ell\ell}}.

\end{document}